\documentclass[twocolumn,showpacs,preprintnumbers,amsmath,prd,amssymb,superscriptaddress,chaproman]{revtex4}

\usepackage{graphicx}
\usepackage{dcolumn}
\usepackage{bm}
\usepackage{multirow}
\usepackage{color}%

\usepackage[english]{babel}
\usepackage{amsmath,amssymb}
\usepackage{amsfonts}
\usepackage{stmaryrd}
\usepackage{mathrsfs}

\newcommand{\Eq}{Eq.~}

\def\nuebar{{\rm \bar{\nu}_e}}
\def\nuebare{{\rm \bar{\nu}_{e}-e}}
\def\nue{{\rm \nu_e}}
\def\nuee{{\rm \nu_{e}-e}}
\def\s2tw{{\rm sin ^2 \theta_{W}}}

\def\cpkkd{\rm{kg^{-1} keV^{-1} day^{-1}}}

\def\er{{\rm \varepsilon_{ee}^{eR}}}
\def\el{{\rm \varepsilon_{ee}^{eL}}}
\def\elr{{\rm \varepsilon_{ee}^{eL,R}}}
\def\etl{{\rm \varepsilon_{e\tau}^{eL}}}
\def\etr{{\rm \varepsilon_{e\tau}^{eR}}}
\def\etlr{{\rm \varepsilon_{e\tau}^{eL,R}}}
\def\eml{{\rm \varepsilon_{e\mu}^{eL}}}
\def\emr{{\rm \varepsilon_{e\mu}^{eR}}}
\def\emlr{{\rm \varepsilon_{e\mu}^{eL,R}}}

\def\Lambdaup{\Lambda_{\cal U}}
\def\dsca{d_{\cal S}}
\def\dvec{d_{\cal V}}

\begin{document}

\preprint{AS-TEXONO/10-03}

\title{Constraints on Non-Standard Neutrino Interactions
and Unparticle Physics\\
with $\nuebar - {\rm e}^-$ Scattering
at the Kuo-Sheng Nuclear Power Reactor}

\newcommand{\as}{Institute of Physics, Academia Sinica, Taipei 11529, Taiwan.}
\newcommand{\metu}{Department of Physics,
Middle East Technical University, Ankara 06531, Turkey.}
\newcommand{\thu}{Department of Engineering Physics, Tsinghua University,
Beijing 100084, China.}
\newcommand{\ihep}{Institute of High Energy Physics,
Chinese Academy of Science, Beijing 100039, China.}
\newcommand{\ciae}{Department of Nuclear Physics,
Institute of Atomic Energy, Beijing 102413, China.}
\newcommand{\bhu}{Department of Physics, Banaras Hindu University,
Varanasi 221005, India.}
\newcommand{\corr}{htwong@phys.sinica.edu.tw;
Tel:+886-2-2789-9682; FAX:+886-2-2788-9828.}

\affiliation{ \as }
\affiliation{ \metu }
\affiliation{ \ihep }
\affiliation{ \thu }
\affiliation{ \ciae }
\affiliation{ \bhu }

\author{ M.~Deniz }  \affiliation{ \as } \affiliation{ \metu }
\author{ S.~Bilmi\c{s}}  \affiliation{ \as } \affiliation{ \metu }
\author{ \.{I}.O.~Y{\i}ld{\i}r{\i}m}  \affiliation{ \as } \affiliation{ \metu }
\author{ H.B.~Li }  \affiliation{ \as }
\author{ J.~Li }  \affiliation{ \ihep } \affiliation{ \thu }
\author{ H.Y.~Liao }  \affiliation{ \as }
\author{ C.W.~Lin }  \affiliation{ \as }
\author{ S.T.~Lin }  \affiliation{ \as }
\author{ M.~Serin } \affiliation{ \metu }
\author{ V.~Singh }  \affiliation{ \as } \affiliation{ \bhu}
\author{ H.T.~Wong } \altaffiliation[Corresponding Author: ]{ \corr } \affiliation{ \as }
\author{ S.C.~Wu }  \affiliation{ \as }
\author{ Q.~Yue }  \affiliation{ \thu }
\author{ M.~Zeyrek } \affiliation{ \metu }
\author{ Z.Y.~Zhou }  \affiliation{ \ciae }

\collaboration{TEXONO Collaboration}  \noaffiliation

\date{\today}

\begin{abstract}

Neutrino-electron scatterings are purely leptonic
processes with robust Standard Model (SM) predictions.
Their measurements
can therefore provide constraints to physics beyond SM.
The $\nuebare$ data
taken at the Kuo-Sheng Reactor Neutrino Laboratory
were used to probe two scenarios:
Non-Standard Neutrino Interactions (NSI) and
Unparticle Physics.
New constraints were placed to the NSI
parameters ($\el$,$\er$),  ($\eml$,$\emr$)
and ($\etl$,$\etr$) for
the Non-Universal and Flavor-Changing channels,
respectively, as well as to the coupling constants 
for scalar ($\lambda_0$) and vector
($\lambda_1$) unparticles to the neutrinos and electrons.

\end{abstract}

\medskip


\maketitle

\section{Introduction}

The compelling evidence of neutrino oscillations
from the solar, atmospheric as well as long baseline
accelerator and reactor neutrino measurements
implies finite neutrino masses and mixings~\cite{pdg08numix}.
Their physical origin and experimental consequences
have not been fully understood yet.
Experimental studies on the neutrino properties
and interactions are crucial because
they can shed light to these
fundamental questions and
may provide hints or constraints to
models on new physics.
Reactor neutrino is an excellent 
neutrino source to address many of
the issues, because of its high flux
and availability. 
The reactor $\nuebar$ spectra is understood
and known, while reactor ON/OFF comparison
provides model-independent means of
background subtraction.

Neutrino-electron scatterings are purely
leptonic processes with robust Standard Model (SM)
predictions~\cite{pdg08sm}.
Experiments on $\nue (\nuebar)$
scattering~\cite{nuephys} have played important
roles in testing SM, and
in the studies of neutrino intrinsic
properties and oscillation.
We report in this paper experimental constraints
on neutrino non-standard interactions (NSI)
and on neutrino unparticle physics (UP) couplings
derived from published results~\cite{texononue,texonomunu,texonocdm}
from $\nuebare$ scattering experiments
at the Kuo-Sheng Nuclear Power Station
in Taiwan.

\section{Electron Antineutrino-Electron Scattering}

\subsection{Standard Model}
\label{sect::nuesm}

The SM cross-section at the laboratory frame
$\nu_{\mu} ( \bar{\nu}_{\mu} ) -$e elastic scattering, where only
neutral-current is involved, is given by~\cite{nuephys,pdg08sm}:
\begin{eqnarray}
\left[ \frac{d\sigma}{dT} ( ^{[-]} \hspace*{-0.35cm} {\nu}_{\mu} e)
\right] _{SM} & = &  \frac{G_{F}^{2}m_{e}}{2\pi }  \cdot
[ ~ \left(g_{V} \pm g_{A} \right) ^{2}  \nonumber \\
& + &  \left( g_{V} \mp g_{A} \right) ^{2}\left(1-
\frac{T}{E_{\nu }}\right) ^{2}  \nonumber  \\
& - & ( g_{V}^2 - g_{A}^2 ) ~ \frac{m_{e}T}{E_{\nu}^{2}} ~ ] ~~~ ,
\label{eq::cs_numu}
\end{eqnarray}
where $G_F$ is the Fermi coupling constant, $T$ is the kinetic
energy of the recoil electron, $E_{\nu }$ is the incident neutrino
energy and $g_{V}$, $g_{A}$ are the vector and
axial-vector coupling constants, respectively. 
The upper(lower) sign refers to the
interactions with $\nu_{\mu} ( \bar{\nu}_{\mu} )$.
The SM assignments to the coupling constants are:
$g_{V}=-\frac{1}{2}+2\s2tw$ and
$g_{A}=-\frac{1}{2}$,
where $\s2tw$ is the weak mixing angle.

The $\rm{ \nu_{e} ( \nuebar ) - e }$
interaction is among the few SM processes which
proceed via {\it both} charged- and neutral-currents,
in addition to their interference effects~\cite{kayser79}.
The cross-section can
be obtained by making
the replacement of
$g_{V,A} \rightarrow ( g_{V,A} + 1 )$
in Eq.~\ref{eq::cs_numu}.
In the case of $\nuebare $ which is relevant for reactor
neutrinos,
\begin{eqnarray}
\left[ \frac{d\sigma}{dT}(\bar{\nu}_{e}e ) \right] _{SM} & = &
\frac{G_{F}^{2}m_{e}}{2\pi }  \cdot
[ ~ \left(g_{V}-g_{A}\right) ^{2}  \nonumber \\
& + & \left( g_{V}+g_{A}+2\right) ^{2}\left(1-
\frac{T}{E_{\nu }}\right) ^{2}  \nonumber  \\
& - & (g_{V}-g_{A})(g_{V}+g_{A} +2)\frac{m_{e}T} {E_{\nu}^{2}}  ~ ]
. \label{eq::gvga}
\end{eqnarray}

By defining chiral couplings $g_L$ and $g_R $:
\begin{eqnarray}
g_L &=& \frac{1}{2}(g_V+g_A) = -\frac{1}{2}+\s2tw
\text{\ \ \ \ and \ \ \ \ } \nonumber \\
g_R &=& \frac{1}{2}(g_V-g_A) = \s2tw \label{eq::glr}~~~ ,
\end{eqnarray}
\Eq \ref{eq::gvga} can be expressed as
\begin{eqnarray}
\left[ \frac{d\sigma}{dT}(\bar{\nu}_{e}e ) \right] _{SM}  &=&
\frac{2G_{F}^{2}m_{e}}{\pi } \cdot [g_{R}^{2} + (g_{L}+1)^{2}(1- \frac{T}{E_{\nu }}) ^{2}  \nonumber  \\
&-& g_{R}(g_{L}+1)\frac{m_{e}T} {E_{\nu}^{2}} ~ ]~~~.
\label{eq::glgr}
\end{eqnarray}

\subsection{Non-Standard Neutrino Interactions}
\label{sect::nsiphys}

Non-standard interactions (NSI) of neutrinos
were introduced in the early work on
neutrino matter effects 
via alternative mechanisms~\cite{earlynsi}.
Models on massive neutrinos generally give rise to NSI.
Examples~\cite{nsimodels} include seesaw type models, low
energy SUSY with R-parity breaking, 
models acquiring mass
radiatively due to the presence of extra Higgs boson, unified SUSY
models as a renormalization effect.
Constraints or evidence of NSI are relevant to the
interpretations of sub-leading contributions
in the forthcoming precision neutrino 
oscillation experiments~\cite{pdg08numix,nsinuosc},
and have consequences in
astrophysics~\cite{nsiastrophys} such as
the understanding of supernova explosion.

\begin{figure}
\includegraphics[width=6cm]{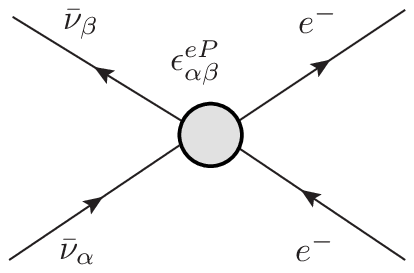} \\[2ex]
\includegraphics[width=7cm]{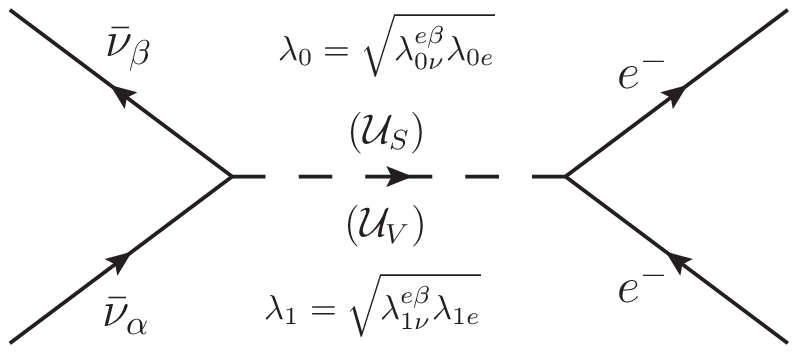}
\caption{ \label{feydiag} 
Top: (a) NSI of neutrinos, generically
described as four-Fermi interaction with new couplings. Bottom:
(b) Interactions of neutrino with electron via exchange of virtual
scalar ${\cal U_S}$ and vector ${\cal U_V}$ unparticle. }
\end{figure}

Phenomenology of NSI
has been explored with
a variety of neutrino sources and interaction 
channels~\cite{nsiastrophys,reactornubsm,nsiboundlsnd,
nsiboundcombined,nsimuon,nsinuN}.
This can be studied with short baseline experiments
where the neutrino fluxes are high and 
the oscillation effects can be neglected.
A model independent approach is
to incorporate
new NSI couplings in the neutrino sector
to the SM electroweak parameters,
as illustrated schematically in Figure~\ref{feydiag}a.
The NSI of $\bar{\nu}_{\alpha} -  e$ scattering is
described by an effective Lagrangian:
\begin{equation}
\mathscr{L}_\text{eff}= -\epsilon_{\alpha\beta}^{eP}
~ 2 \sqrt{2}G_F
(\bar{\nu}_{\alpha}\gamma_{\rho}L\nu_{\beta})(\bar{e}\gamma^{\rho}Pe) ~~ ,
\end{equation}
where $\epsilon_{\alpha\beta}^{eP}$ describes the coupling strength
with respect to $G_F$. 
The helicity states are denoted by P (=L,R),
and  $(\alpha , \beta)$ stand for 
the lepton flavor (e, $\mu$ or $\tau$). 
The cases where 
$\alpha = \beta$ and $\alpha \ne \beta$ 
correspond to Non-Universal (NU)
and Flavor-Changing (FC) NSI, respectively.

For reactor neutrinos, $\alpha = e$, and
six parameters are involved $-$
the NU $\elr$ as well as the FC  $\emlr$ and $\etlr$.
The cross-section formula including 
both SM and NSI interactions for
$\nuebar + e \rightarrow \nuebar + e$
is given by~\cite{nsiboundcombined,nsiboundlsnd}
\begin{eqnarray}
\left[\frac{d\sigma}{dT}\right]_{SM+NSI} = ~~\frac{2 G_F^2 m_e}{\pi}
\cdot [~\left(\tilde g_R^2 + \sum_{\alpha \neq e}|\epsilon_{\alpha
e}^{e R}|^2 \right)\nonumber \\
+ \left((\tilde g_L + 1)^2 + \sum_{\alpha \neq e}|\epsilon_{\alpha
e}^{e L}|^2 \right) \left(1 - \frac{T}{E_{\nu}}\right)^2\nonumber \\
- \left(\tilde g_R(\tilde g_L + 1)+\sum_{\alpha \neq
e}|\epsilon_{\alpha e}^{e R}||\epsilon_{\alpha e}^{e L}|
\right)\frac{m_e T}{E^2_{\nu}}] ~~~ ,
\label{nsics}
\end{eqnarray}
where $\tilde g_L=g_L+\el$ and $\tilde g_R=g_R+\er$.

\begin{figure}
\includegraphics[width=8cm]{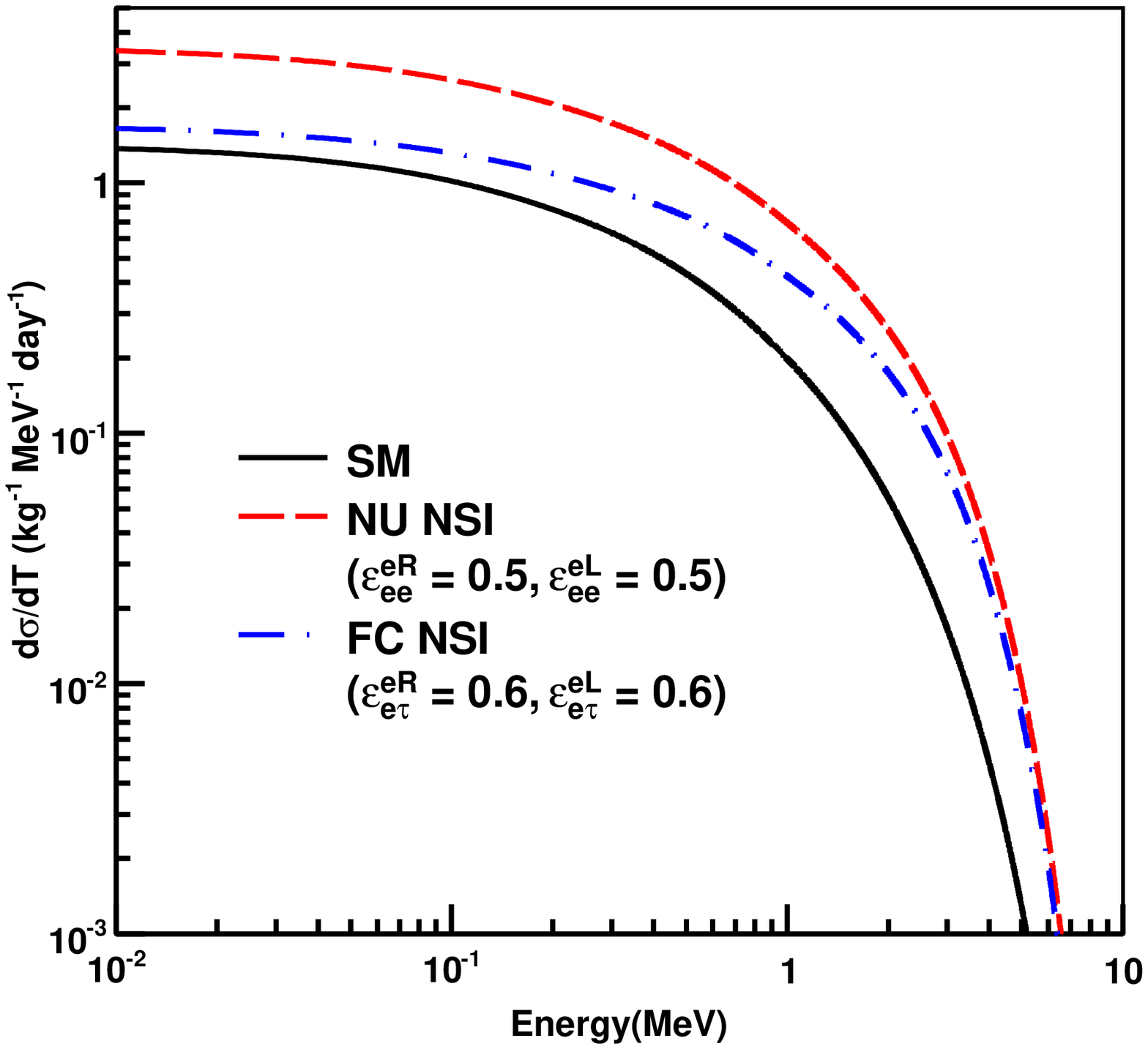} \\[2ex]
\includegraphics[width=8cm]{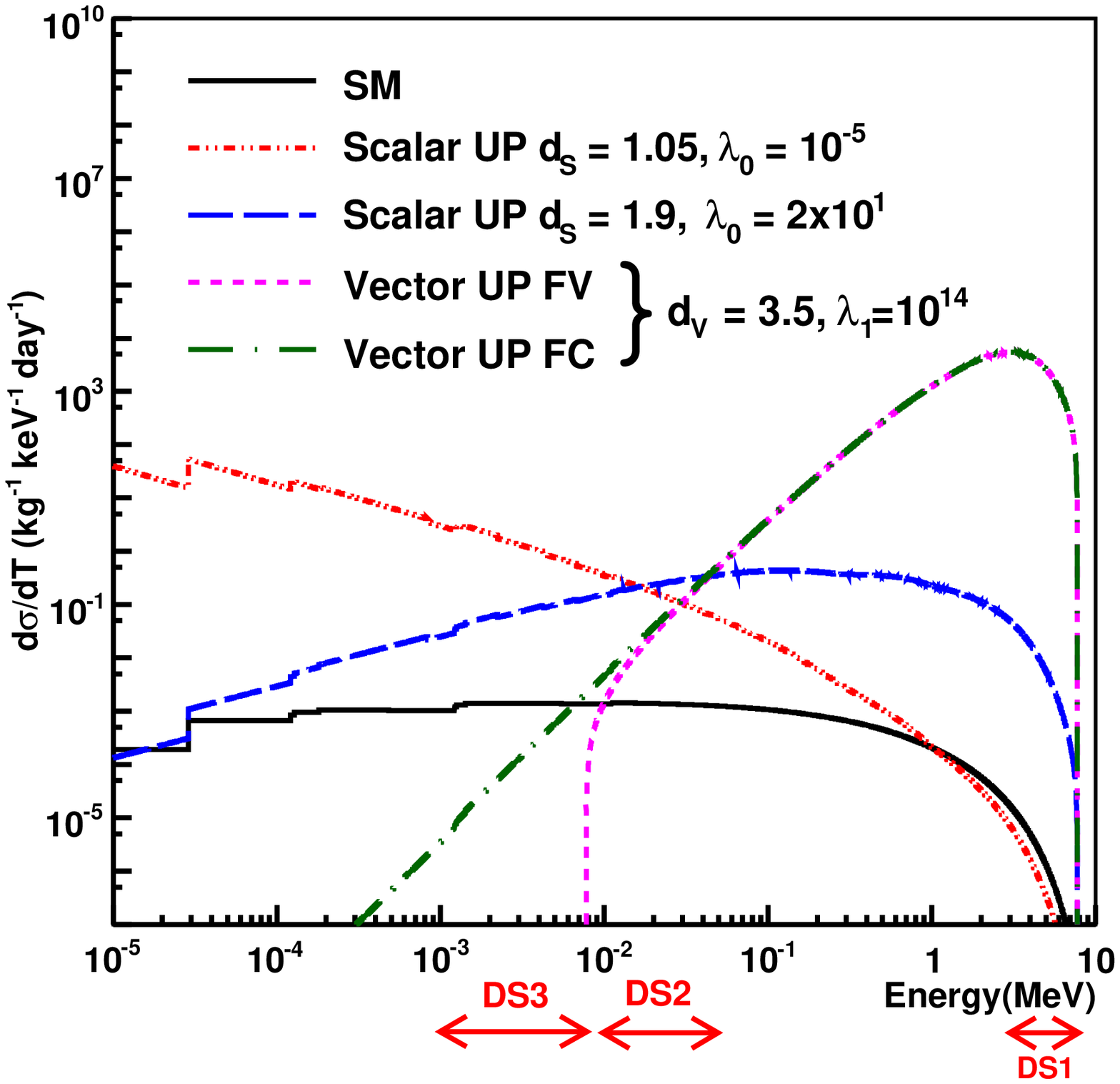}
\caption{ \label{dsdT} 
Differential cross-section as function of the
recoil energy $T$ with typical reactor-$\nuebar$ spectra.
Top: (a) NSI at coupling
parameters relevant to this work
using CsI(Tl) as target. 
Bottom: (b) scalar UP, at two values of 
$( \dsca , \lambda_{0} )$
and vector UP, at a value of 
$( \dvec , \lambda_{1} )$ for both FV and FC cases,
using Ge as target. 
The SM contributions are also superimposed.
The relevant energy ranges of the three data sets 
used in the present analysis are also shown.
}
\end{figure}

The measurable recoil spectra at a typical reactor
flux of $\phi (\nuebar) = 10^{13} ~ {\rm cm^{-2} s^{-1}}$ are
displayed in Figure~\ref{dsdT}a, at NSI parameters
in both NU and FC channels relevant to this work. 
The SM spectrum is superimposed. The NSI
contributions give rise to similar spectral shapes as the SM one.
Accordingly, the appropriate strategy to study NSI is to focus at
the MeV energy range where the SM effects were measured with good
accuracy~\cite{texononue}.

The strong experimental limits on the branching ratio of 
$\mu \rightarrow 3 e$ in accelerator experiments
provided stringent bounds 
on $|\emlr|<5\times10^{-4}$~\cite{nsiboundlsnd},
which is highly sensitive to loop processes.
Model independent analysis after taking gauge
invariance into account gave rise to
weaker bounds on 
$|\emlr| < 0.1 $~\cite{nsimuon}.
We present results on
the FC parameters 
$\emlr$ and $\etlr$, as well as 
the NU parameters $\elr$
in this analysis with reactor $\nuebar$ data.

\subsection{Unparticle Physics}

A scale invariant sector can be described by Banks-Zaks (BZ) fields
which is related to gauge theories with non-integer number of
fermions~\cite{Banks1981}.
BZ fields has its own gauge group and do
not couple to the SM fields which have definite masses.
It has been proposed~\cite{Georgi2007} that both the SM
and BZ fields may coexist in a high energy scale.
Below an energy scale $\Lambdaup$, BZ
operators turn into unparticle operators
${\cal O_U}$ with a non-integer scaling
dimension, denoted by $\dsca$ and $\dvec$
for the scalar and vector cases, respectively.

Unparticle effects can be studied
in accelerator experiments~\cite{upcollider} through
their direct production,
the signatures of which are
missing energy in the detectors.
An alternative method is to probe the
virtual effects of unparticles which act as
mediators in the interactions~\cite{Georgi2007,upcollider}.
This approach was adopted in the present analysis 
using reactor neutrinos as probe.
The interaction Lagrangians
for $\nu_{\alpha} + e \rightarrow \nu_{\beta} + e$
via virtual scalar and vector unparticle exchange
as depicted in Figure~\ref{feydiag}b
are given, respectively, 
by~\cite{upcollider,upchen07,upbalantekin07,upbarranco09}
\begin{eqnarray}
\mathscr{L}_\text{J=0}  & = &
\frac{\lambda_{0 e}}{\Lambdaup^{\dsca - 1}}\,\bar e e \,{\cal O_U} +
\frac{\lambda_{0\nu}^{\alpha\beta}}{\Lambdaup^{\dsca - 1}} \,
\bar \nu_\alpha \nu_\beta \,{\cal O_U}  ~~ {\rm and} \\
\mathscr{L}_\text{J=1} & = &
\frac{\lambda_{1 e}}{\Lambdaup^{\dvec - 1}}\,\bar e \gamma_\mu e \,
{\cal O}_{\cal U}^\mu +
\frac{\lambda_{1\nu}^{\alpha\beta}}{\Lambdaup^{\dvec - 1}} \,
\bar \nu_\alpha \gamma_\mu \nu_\beta \,{\cal O}_{\cal U}^\mu ~ ,
\end{eqnarray}
where $\lambda_{J e}$ and $\lambda_{J \nu}^{\alpha \beta}$ are the
corresponding coupling constants with $J = 0,1$ denoting
scalar and vector unparticle interactions, respectively.

The cross-section of $\bar{\nu}_{e} -$e scattering with
scalar unparticle exchange is given by
\begin{equation}
\label{cs:e-scalar}
\left(\frac{d\sigma}{dT}\right)_{\cal U_S}
=\frac{f_0^2(\dsca)}{\Lambdaup^{4\dsca-4}}~
\frac{2^{2\dsca-6}}{\pi E_\nu^2}~(m_eT)^{2d-3}~(T+2m_e) ~~,
\end{equation}
where
\begin{equation}
f_0 (\dsca)=\frac{\lambda_{0 \nu}^{\alpha\beta}\lambda_{0 e}}{2\sin(\dsca\pi)}
A_0 ( \dsca )
\label{eq::f}
\end{equation}
and the normalization constant
$A_0 (\dsca) $ is given by:
\begin{equation}
A_0 ( \dsca ) = \frac{16\pi^{5/2}}{(2\pi)^{2\dsca}}
\frac{\Gamma(\dsca+1/2)}{\Gamma(\dsca-1)\Gamma(2\dsca)} ~~.
\label{eq::a}
\end{equation}
The interference effects with SM 
are negligible due to suppression by
factors of  $m_\nu/\Lambdaup$. 
Therefore, it is not necessary to differentiate
flavor conserving (FC) and violating (FV)
scalar UP interactions.

The cross-section of $\nuebare$ scattering
via vector UP exchange is
\begin{eqnarray}
\left(\frac{d\sigma}{dT}\right)_{\cal U_V}
&=& \frac{1}{\pi}~
\frac{f_1^2 (\dvec)}{\Lambdaup^{4\dvec-4}}
~2^{2\dvec-5}~m_e^{2\dvec-3}~T^{2\dvec-4}\nonumber\\
&\times& \left[1 + \left(1-\frac{T}{E_\nu} \right)^2
-\frac{m_eT}{E_\nu^2}\right] ~~ ,
\label{cs:e-vector}
\end{eqnarray}
where $f_1 (\dvec)$ follows a similar
expression as Eq.~\ref{eq::f},
making the replacement 
$ \lambda_{0 \nu}^{\alpha\beta}\lambda_{0 e}
\rightarrow \lambda_{1 \nu}^{\alpha\beta}\lambda_{1 e}$
and 
$A_0 ( \dsca ) \rightarrow A_1 ( \dvec )$.
Unlike the scalar UP case,
the interference effects with SM  
also contribute in the vector UP interactions:
\begin{eqnarray}
\left(\frac{d\sigma}{dT}\right)_{\cal U_V-SM}&=& \frac{\sqrt{2}G_F}{\pi}~
\frac{f_1(\dvec)}{\Lambdaup^{2\dvec-2}}~(2m_e T)^{\dvec-2}~m_e \nonumber\\
&\times&[ g_R + (g_L+1) \left(1-\frac{T}{E_\nu}\right)^2 \nonumber\\
&-& \frac{\left(g_L+g_R+1\right)}{2} \frac{m_e T}{E_\nu^2}] ~ .
\label{cs:e-vector-interf}
\end{eqnarray}
The FV and FC cross-sections for vector UP 
are therefore given by
Eq.~\ref{cs:e-vector} and
the sum of Eq.~\ref{cs:e-vector} and Eq.~\ref{cs:e-vector-interf},
respectively.

The differential cross-sections of 
the UP interactions
using Ge as target
are displayed in Figure~\ref{dsdT}b with the SM contributions
superimposed for comparison. 
The saw-tooth structures 
for $T \alt 1 ~{\rm keV}$ are due to suppression
by the atomic binding energy~\cite{atombinding}.

Three sets of parameters characterize the unparticle 
interactions and can be probed experimentally: 
(i) unparticle energy scale $\Lambdaup$, 
(ii) unparticle mass dimensions $\dsca$ and $\dvec$, 
as well as (ii) coupling constants 
$ \lambda_0 \equiv \sqrt{ \lambda ^ {e\beta}_{0\nu} \lambda_{0e}}$ and 
$ \lambda_1 \equiv \sqrt {\lambda^{e\beta}_{1\nu} \lambda_{1e}} $ 
for the scalar and vector UP interactions, respectively. 
The UP energy scale is taken to be 
$\Lambdaup \sim 1~ {\rm TeV}$ in most
recent work~\cite{upbalantekin07,upmontanino08,upbarranco09}.
Unitarity requirement placed constraints on the
dimension~\cite{updimension} to be within $1< \dsca <2$ in the
scalar case, but only provides lower bound $\dvec \geq 3$  for
vector UP exchange.
The spectral shape of Figure~\ref{dsdT}b and the $T$-dependence of
Eq.~\ref{cs:e-scalar} indicate that measurements with low energy
threshold are expected to provide better sensitivities at $\dsca  <
3/2$. On the other hand, high energy experiments are
preferred to probe UP due to the large values of $\dsca$
and $\dvec$.

\section{Experimental Constraints}

\subsection{Merits of Reactor Neutrinos}

Several features make short-baseline reactor neutrino experiments
optimal for probing physics beyond 
SM~\cite{reactornubsm}. Reactor neutrinos are pure $\nuebar$ which
simplifies interpretations of the results. Atmospheric and solar
neutrinos have different eigenstate compositions at the detectors.
The constraints from these experiments therefore are not identical
to those at reactors, analogous to the studies of neutrino magnetic
moments~\cite{effbounds}.

Experimentally reactors produce high $\nuebar$-fluxes
compared to other sources. The reactor OFF periods
provide model-independent means of background subtraction.
It was recently recognized~\cite{texononue} that
the studies of reactor $\nuebare$
provide better sensitivities to
the SM electroweak parameters $\s2tw$
and $( g_V , g_A )$
at the same experimental accuracies
as those from $\nue -$e measurements.
The lower neutrino energy at the MeV range also
favors applications where sensitivities
can be enhanced at low detector threshold.

\begin{figure}
\includegraphics[width=7.6cm]{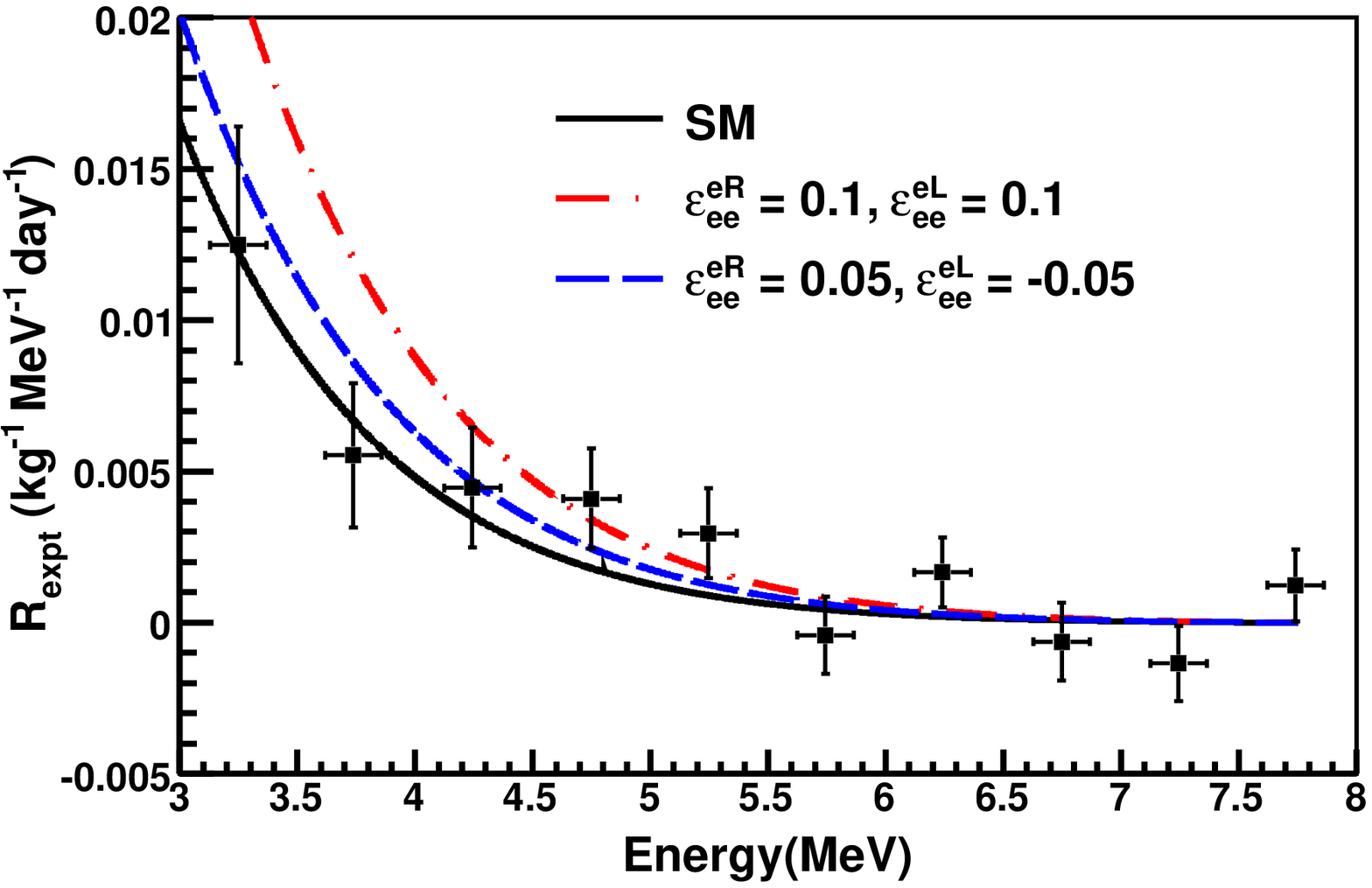} \\[2ex]
\includegraphics[width=7.6cm]{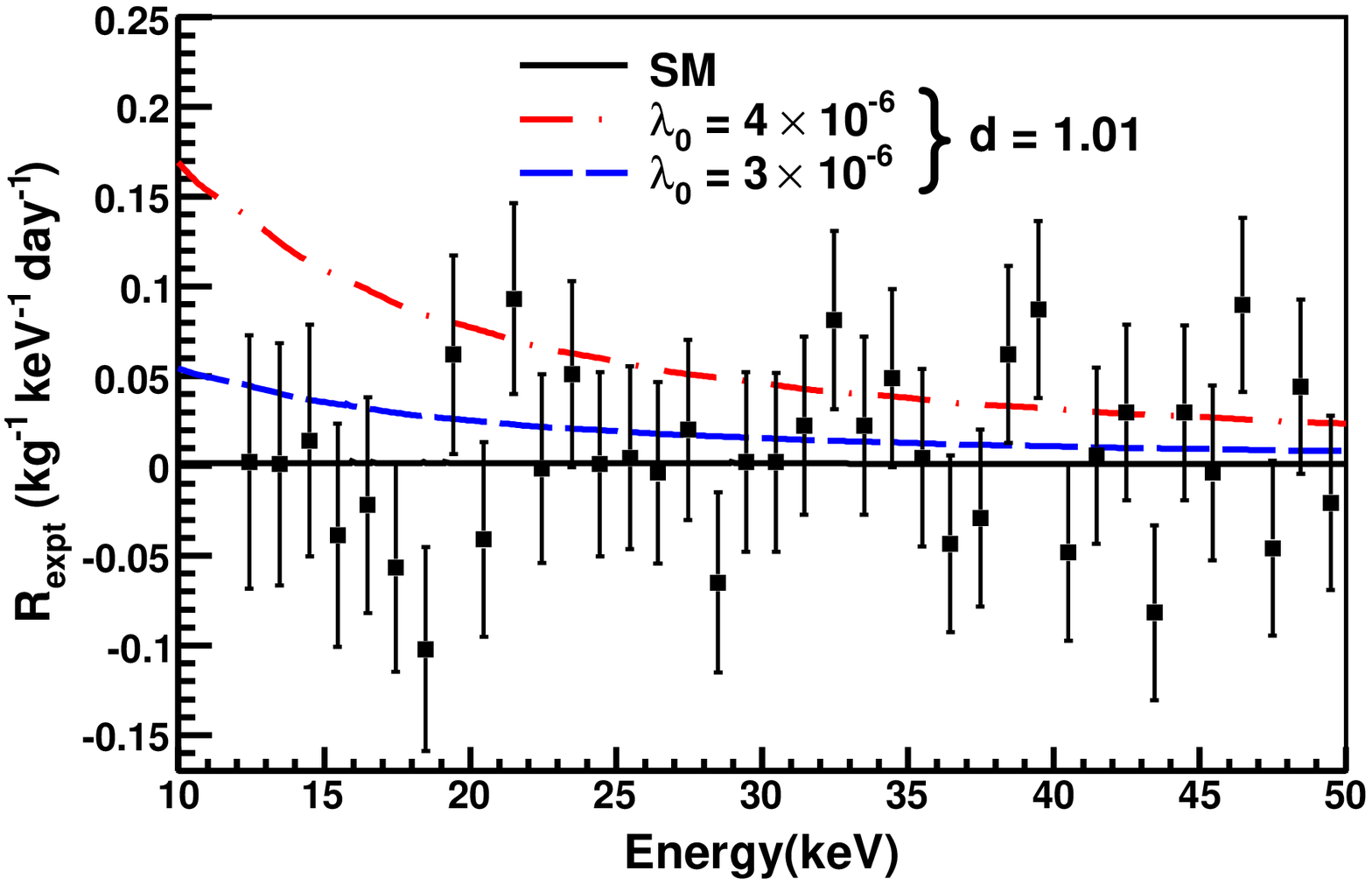} \\[2ex]
\includegraphics[width=7.6cm]{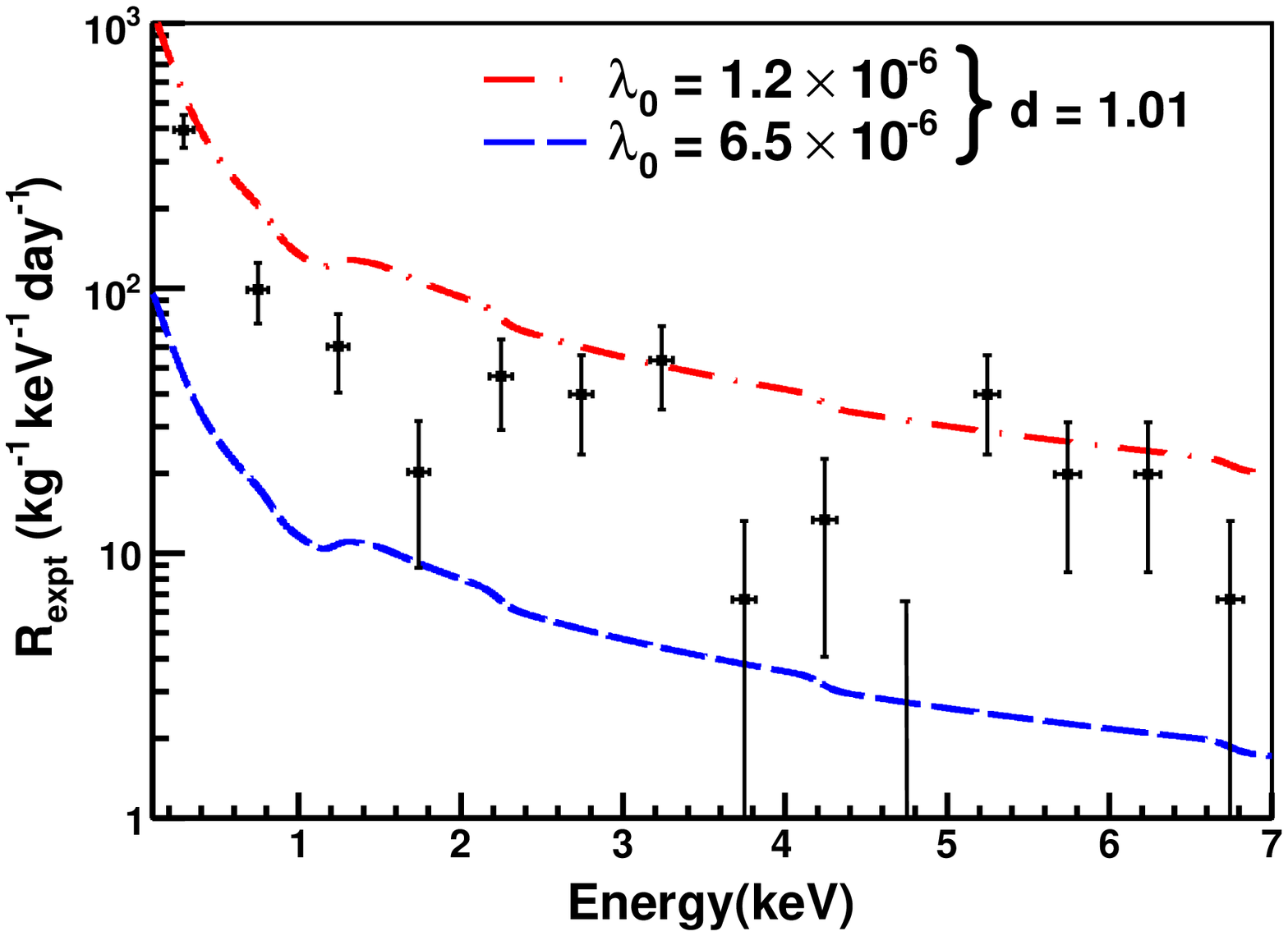}
\caption{ 
\label{fig::data}
The three data sets 
adopted for this analysis.
Observable NSI or UP spectra at 
allowed and excluded parameter space
are superimposed.
Top: (a) DS1-CsI(Tl) Reactor ON$-$OFF~\cite{texonomunu},
showing SM+NSI
with NSI at $( \er , \el )$ = (0.1,0.1)
and (0.05,$-$0.05).
Middle: (b) DS2-HPGe Reactor ON$-$OFF~\cite{texononue}, 
showing SM+UP with 
$\lambda_0 = 4 \times 10^{-6} ~ {\rm versus} ~ 
3 \times 10^{-6}$ at $\dsca = 1.01$.
Bottom: (c) DS3-ULEGe Reactor ON only~\cite{texonocdm}, 
showing SM+UP with
$\lambda_0 = 1.2 \times 10^{-5} ~ {\rm versus} ~
6.5 \times 10^{-6}$ at $\dsca = 1.01$.
The SM contributions from $\nuebar -$e
are displayed in (a) and (b) as comparison,
and are out of range at $\sim 10^{-3} ~ \cpkkd$ 
in (c).
}
\end{figure}

\subsection{Input Data}

\begin{table*} [hbt]
\caption{ 
Constraints at 90\% CL due to one-parameter fits on the
NSI couplings. The results
are presented as 
``best-fit $\pm$ statistical error $\pm$ systematic error''.
Bounds from LSND~\cite{nsiboundlsnd} and
combined data~\cite{nsiboundcombined},
as well as from a model-independent
analysis~\cite{nsimuon}
are compared with those of this work. 
The projected statistical sensitivities 
correspond to a potential measurement of the 
SM cross-section at 2\% accuracy~\cite{texononue}.} 
\label{tab:eelr}
\begin{ruledtabular}
\begin{tabular}{ccccccccccc}
& & \multicolumn{3}{c}{TEXONO (This Work)}  & & 
LSND~\cite{nsiboundlsnd} & 
Combined~\cite{nsiboundcombined} & Ref.~\cite{nsimuon} \\
\multicolumn{2}{c}{NSI} & Measurement & & Bounds & Projected & & & & \\
\multicolumn{2}{c}{Parameters} &  
Best-Fit & $\chi ^2$/dof & at 90\% CL &
Sensitivities &
\multicolumn{3}{c}{Bounds at 90\% CL}  \\ \hline

\multirow{4}*{NU \{} & $\el$
& $\el =$ & 8.9/9 &  $-1.53 < \el < 0.38 $  &
$\pm$0.015 &
$-0.07 < \el < 0.11 $ &
$-0.03 < \el < 0.08 $ & $ | \el | < 0.06$  \\
& & $0.03 \pm 0.26 \pm 0.17$ & & & & & & \\

& $\er$
&  $\er =$ &  8.7/9  & $-0.07 < \er < 0.08 $  &
$\pm$0.002 &
$-1.0 < \er < 0.5 $ & 
$0.004 < \er < 0.151$  & $ | \er | < 0.14 $ \\ 
& &   $0.02 \pm 0.04 \pm 0.02$ & & & & & & \\ \hline

  \multirow{4}*{FC \{} & $\eml$ \multirow{2}*{\{}
& $\eml ^2 ( \etl ^2 ) =$
&  \multirow{2}*{\} 8.9/9}  & $|\eml| < 0.84$   &
$\pm$0.052 &
$-$  & $ |\eml| < 0.13$  &  $ |\eml| < 0.1$ \\

& $\etl$ ~~ 
& $0.05 \pm 0.27 \pm 0.24$ &  & $|\etl| < 0.84$   &
$\pm$0.052 &
$|\etl| < 0.4$ & 
$ |\etl| < 0.33 $ & $ | \etl | < 0.4 $ \\

& $\emr$  \multirow{2}*{\{} 
&  $\emr ^2 ( \etr ^2 ) =$ 
&  \multirow{2}*{\} 8.7/9}  & $ |\emr| < 0.19 $  &
$\pm$0.007 &
$-$  & $ |\emr| < 0.13 $  &  $ |\emr| < 0.1$ \\

& $\etr$ ~~ 
& $0.008 \pm 0.015 \pm 0.012$ &  & $ |\etr| < 0.19 $  &
$\pm$0.007 &
$ |\etr| < 0.7 $ & 
$ 0.05 < |\etr| < 0.28 $ & $ | \etr | < 0.27 $ \\
\end{tabular}
\end{ruledtabular}
\end{table*}

Data adopted for this analysis were taken at
the Kuo Sheng Neutrino Laboratory (KSNL)
located at a distance of 28 m from the reactor core.
The nominal thermal power output was 2.9 GW
producing  an average $\nuebar$-flux of
$\phi ( \nuebar )
\sim {\rm 6.4 \times 10^{12} ~ cm^{-2} s^{-1}}$~\cite{texonomunu}.
Detectors were placed inside a shielding structure
where ambient radioactivity was suppressed
by 50~tons of passive materials.

Three independent data sets were adopted, each having
a different energy range as depicted in Figure~\ref{dsdT}.
\begin{description}
\item[\bf DS1-CsI(Tl):]
Data with 29882/7369 kg-days of Reactor ON/OFF exposure 
of a CsI(Tl) crystal scintillator array~\cite{texononue}
with a total mass of 187~kg. Analysis range is $\rm{3 - 8 ~
MeV}$. 
From the excess of events in the ON$-$OFF residual spectrum,
the SM electroweak angle was measured to be $\s2tw =  0.251
\pm 0.031 (stat) \pm 0.024 (sys)$ which improved over previous
results from $\nuebar -$e scattering and was comparable to those
from $\nue -$e experiments.
\item[\bf DS2-HPGe:]
Data with 570.7/127.8~kg-days of Reactor ON/OFF exposure
taken with a high-purity germanium (HPGe) detector~\cite{texonomunu}
with a target mass of 1.06~kg.
Analysis threshold of 10~keV with a background
level of $\sim 1 ~ \cpkkd$ was achieved.
The low threshold allowed sensitive limits 
on neutrino magnetic moments to be derived
from the ON$-$OFF residual spectrum.
\item[\bf DS3-ULEGe:]
Data with  0.338~kg-days of Reactor ON exposure
taken with an ultra-low-energy germanium (ULEGe) detector
array~\cite{texonocdm} with a total mass of 20~g
and a threshold of 220$\pm$10~eV.
The sub-keV threshold opened a window of studying
WIMP dark matter with mass less than 10~GeV.
\end{description}

The three data sets (DS1$-$3)
are displayed in Figures~\ref{fig::data}a,b\&c,
respectively. 
Their respective energy ranges are 
depicted in Figure~\ref{dsdT}.
The SM contributions from $\nuebar -$e
are superimposed in (a) and (b),
and are out of range at 
$\sim 10^{-3} ~ \cpkkd$ in (c).
The NSI or UP scenarios where the data sets 
would be optimal to provide sensitive bounds were selected. 
The observable spectra of an
excluded and an allowed parameter space were superimposed as
illustrations.

The observed event rates ($R_{expt}$)
of the various data sets,
in units of $\cpkkd$,
were compared to the expected rates ($R_X$) evaluated for
the different interaction channels $X$ ($X=SM,NSI,UP$), via
\begin{equation}
R_{X} ~ = ~ \rho_e ~ \int_{T} \int_{E_{\nu }}
\left(\frac{d\sigma}{dT}\right)_{X}  ~ \frac{d \phi ( \nuebar ) }{dE_{\nu}} ~
dE_{\nu} ~ dT ~~ ,
\label{eq::NPpar}
\end{equation}
where $\rho_e$ is the electron number density per kg of target mass,
and $d \phi ( \nuebar ) / d E_{\nu}$ corresponds to
the neutrino spectrum. Constraints were then derived.

\begin{figure}
\includegraphics[width=8cm]{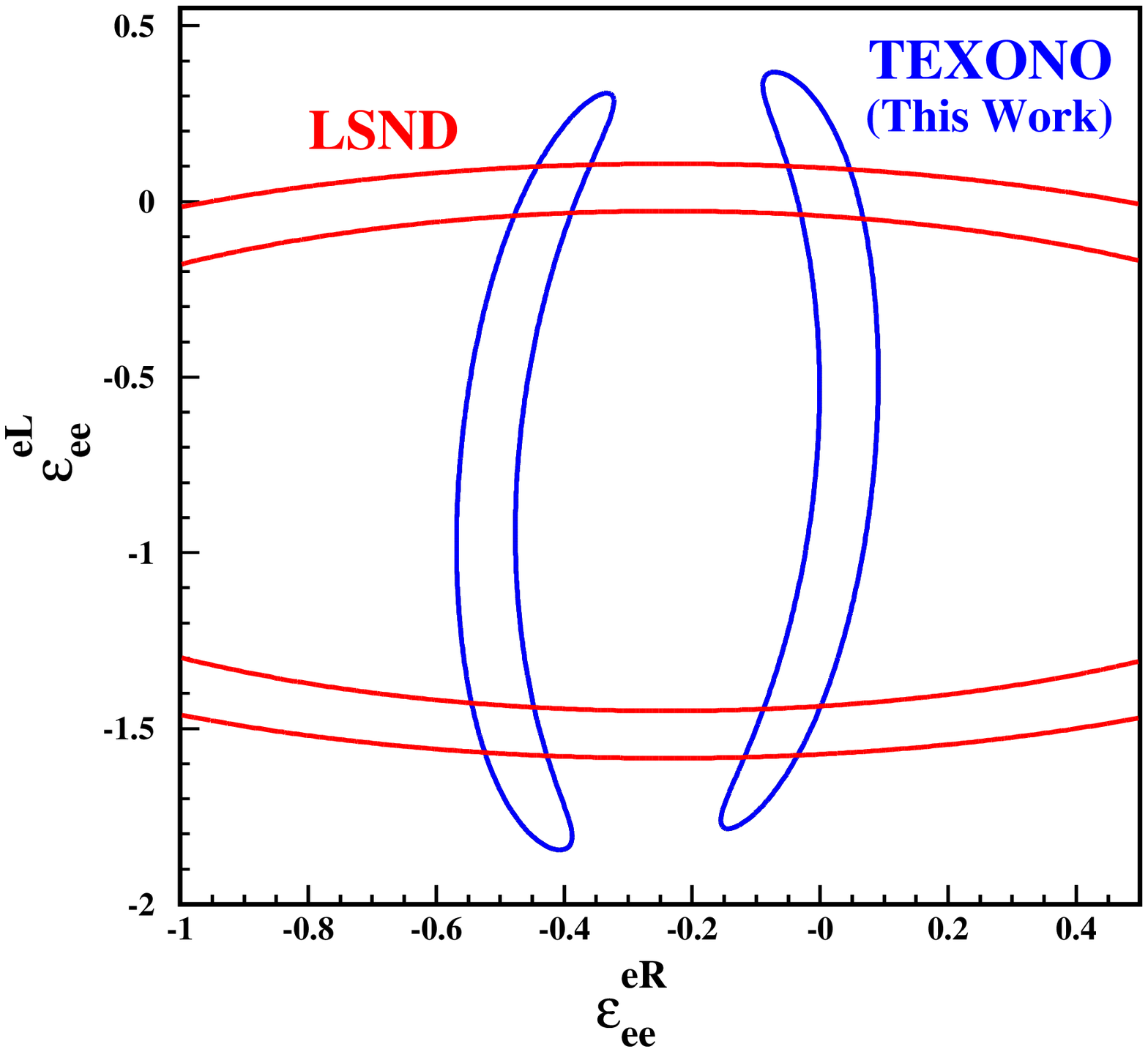} \\[2ex]
\includegraphics[width=8cm]{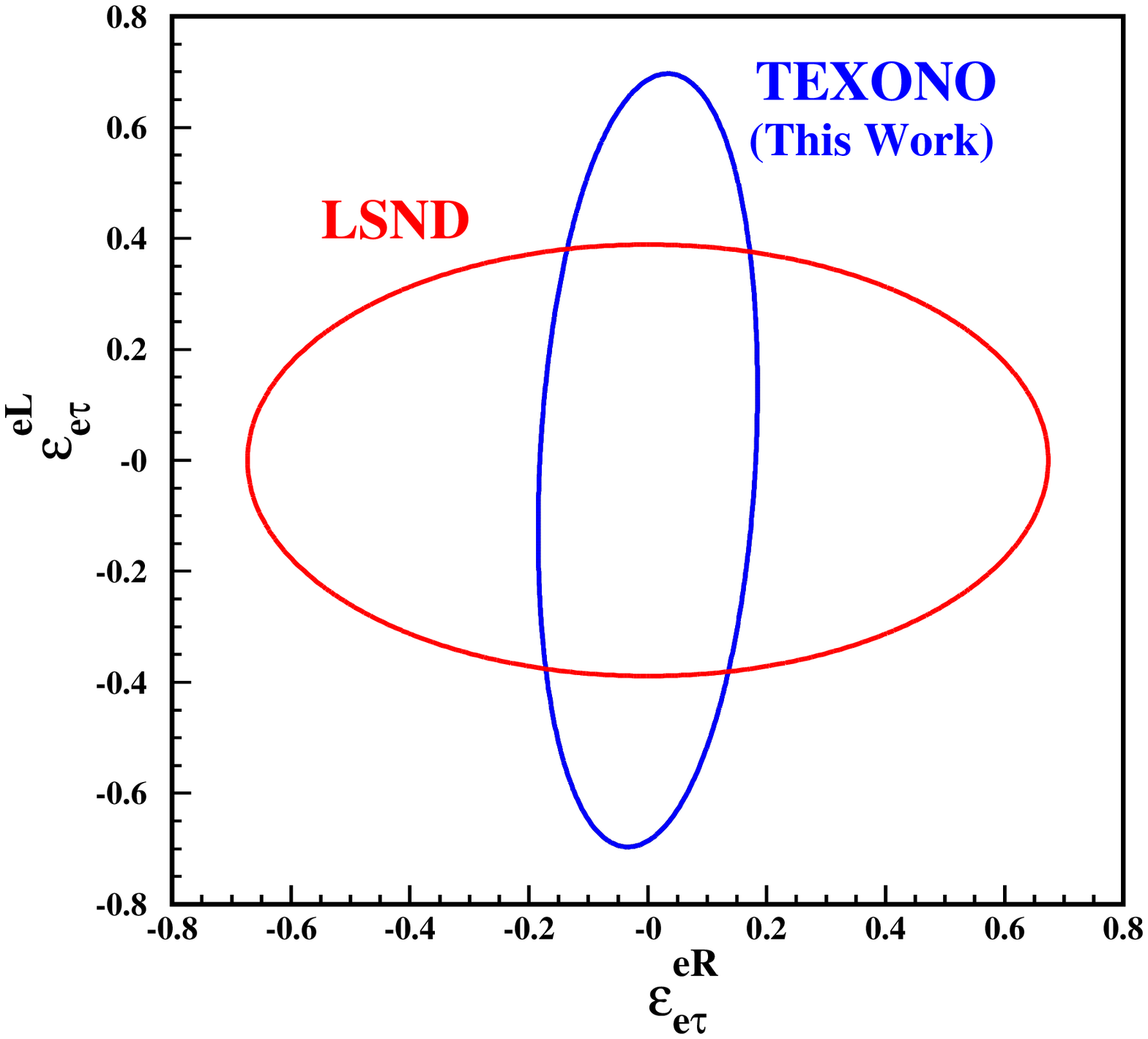}
\caption{ \label{nsitexono}
The allowed region at 90\% CL for
Top: (a) NU NSI parameters of $\el$ and $\er$;
Bottom: (b) FC NSI
parameters of $\etl$ and $\etr$
from DS1-CsI(Tl) on $\nuebar -$e.
The allowed regions from the LSND experiment on 
$\nue -$e
are superimposed.
The constraints in the  $( \eml , \emr )$ plane 
are the same as those of  $( \etl , \etr )$ in (b).
}
\end{figure}

Different analysis algorithms were necessary for the three data
sets. A minimum-$\chi ^2$ fit was performed for 
DS1-CsI(Tl) and DS2-HPGe, 
with
\begin{equation}
\chi ^{2}=\sum_{i=1} \left[ \frac{  R_{expt}(i) -
[ R _{SM} (i) + R_{X} (i) ]}{\Delta_{stat} (i)}
\right]^{2} ~~ ,
\label{eq::chi2UP}
\end{equation}
where $R_{SM} (i)$  and $R_{X} (i)$
are the expected event rates
on the $i^{\rm th}$ data bin
due to the SM and X(=NSI or UP) contributions, respectively,
while $\Delta_{stat} (i) $
is the corresponding uncertainty of the measurement.
For DS3-ULEGe, there was no corresponding Reactor OFF data so that
the conventional Reactor ON$-$OFF background
subtraction  and a best-fit analysis were not possible.
Instead, the ``Binned Poisson'' method
developed for dark matter searches~\cite{cdm09}
was adopted.
No background assumption was made
such that upper bounds on 
NSI or UP-induced contributions
were placed  since
they could not be larger than 
the observed signals.

\begin{figure}
\includegraphics[width=8cm]{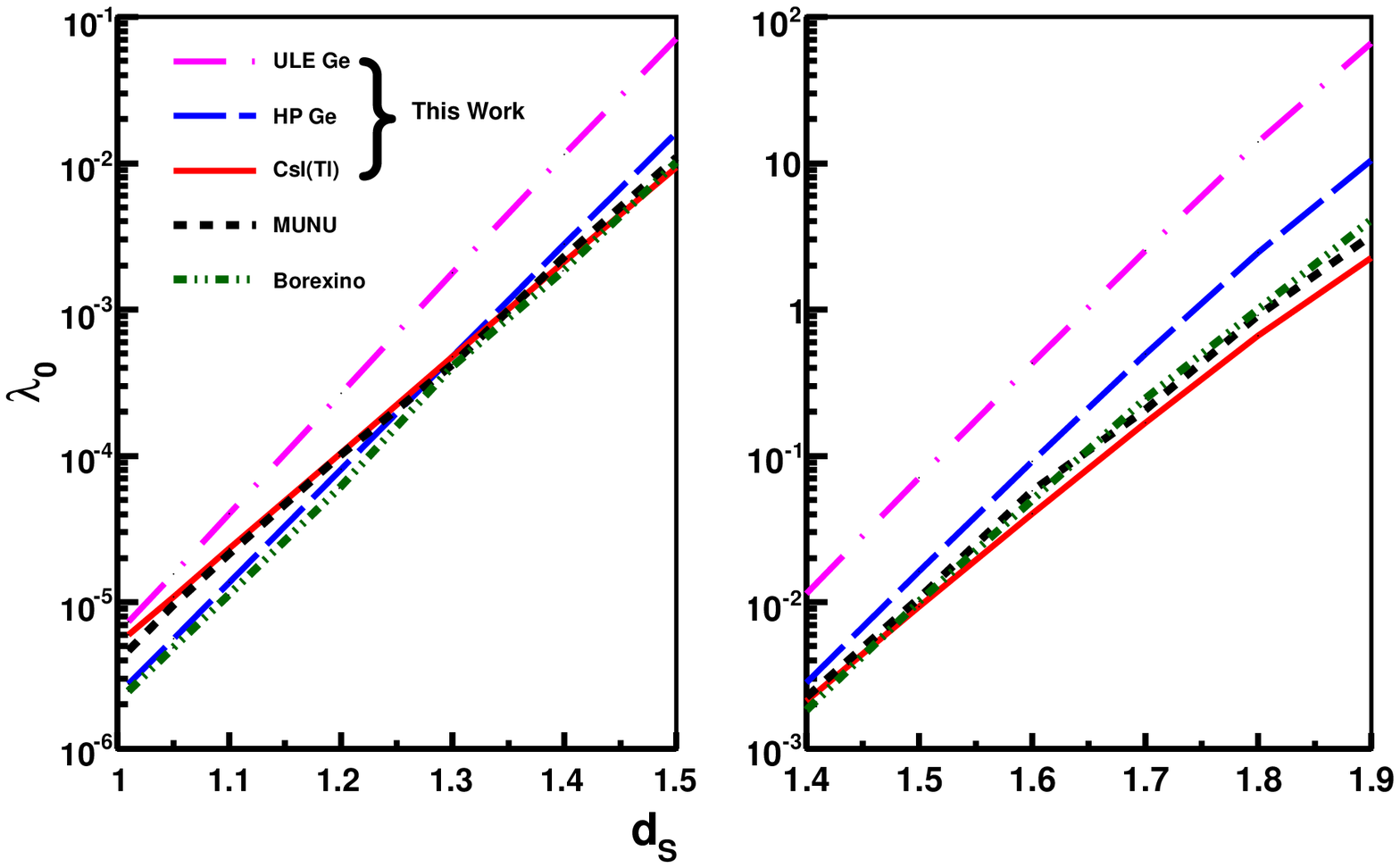} \\[2ex]
\includegraphics[width=8cm]{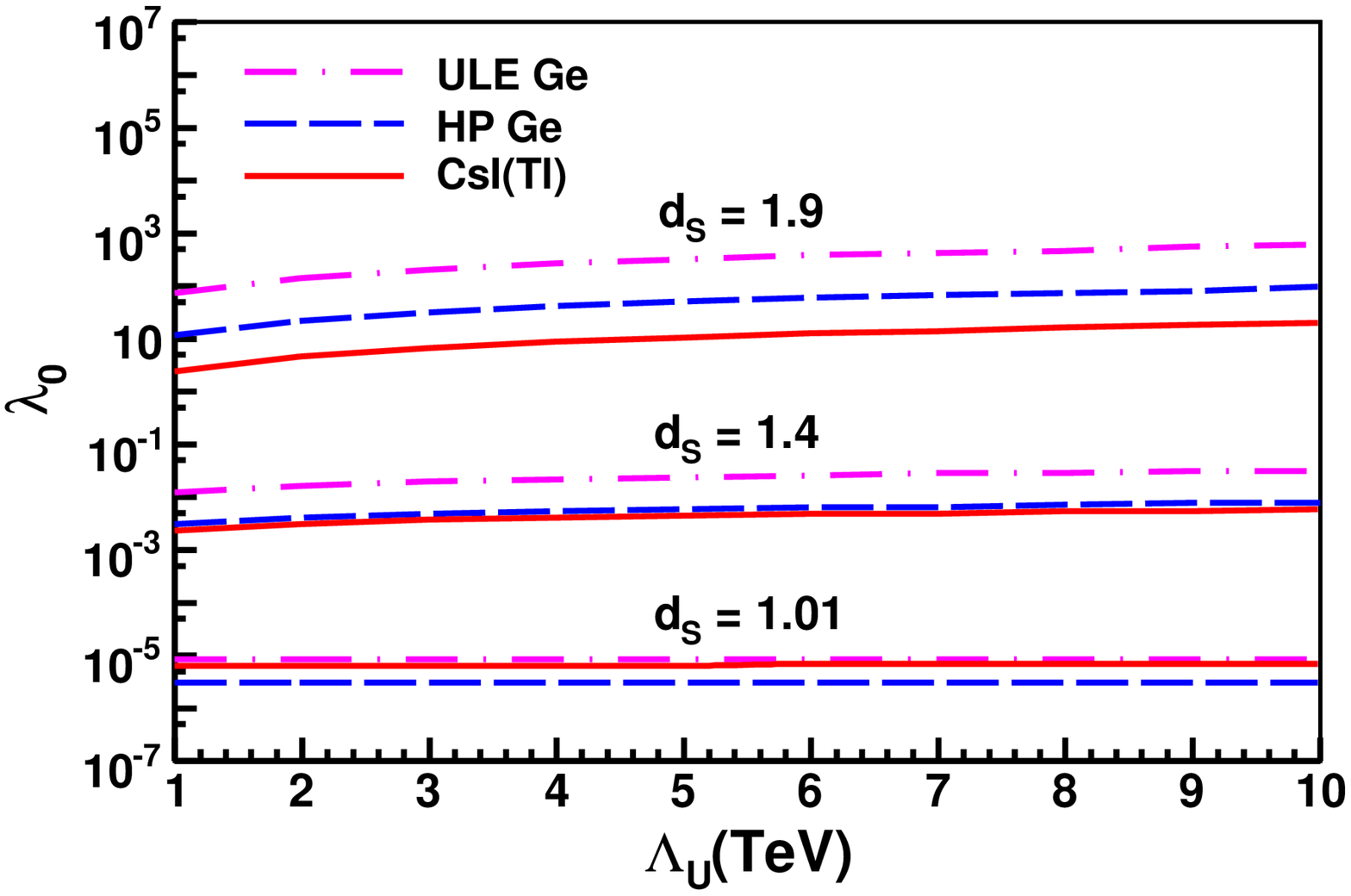}
\caption{ \label{upscalar} Constraints on UP with
scalar exchange $-$ Top: (a) The coupling $\lambda_0$ versus
mass dimension $d_{\cal S}$ at $\Lambdaup = 1 ~ {\rm TeV}$; 
Bottom: (b) 
Upper bounds on $\lambda_0$ at different energy scales
$\Lambdaup$. 
Parameter space above the lines is excluded.
}
\end{figure}

\begin{figure}
\includegraphics[width=8cm]{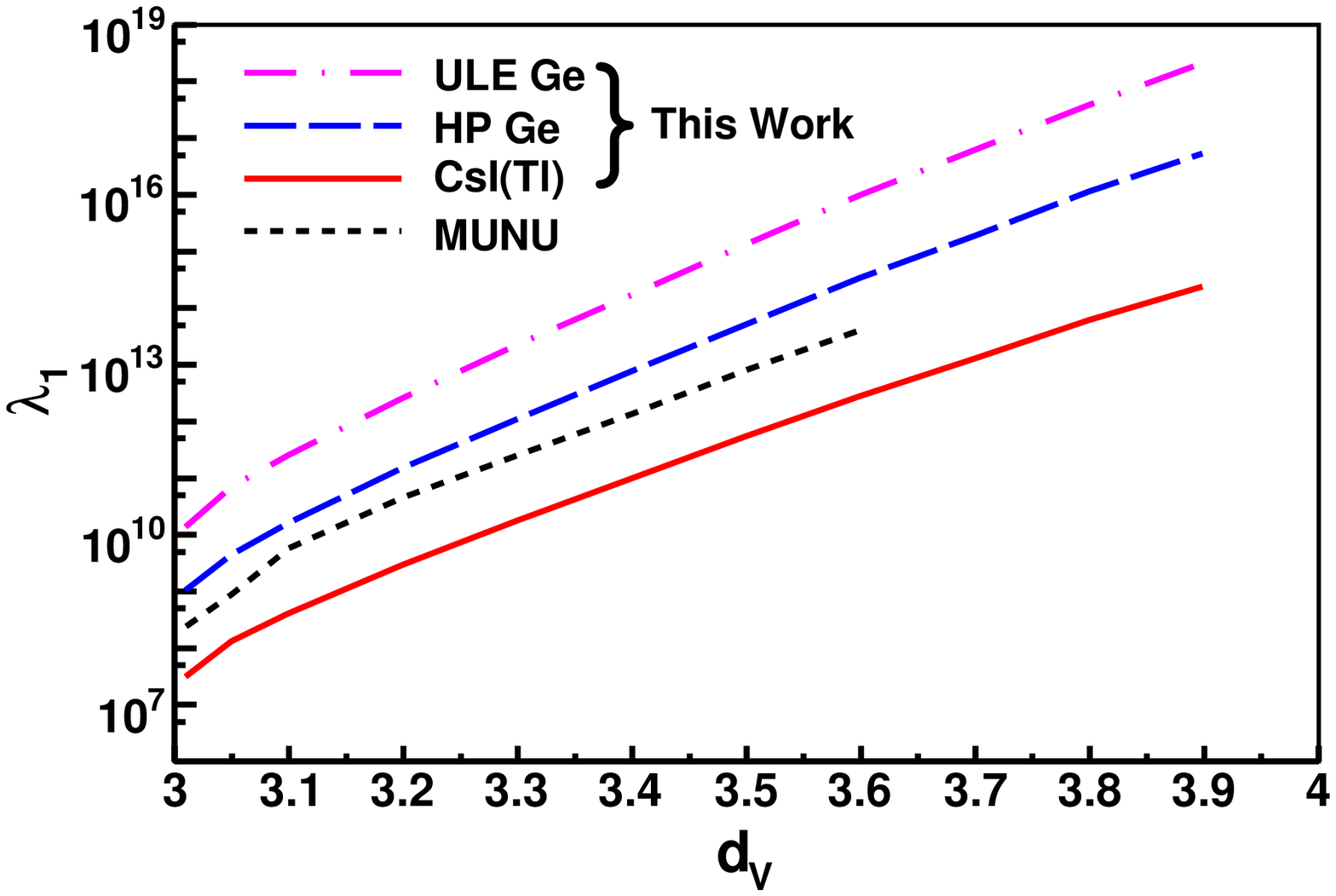} \\[2ex]
\includegraphics[width=8cm]{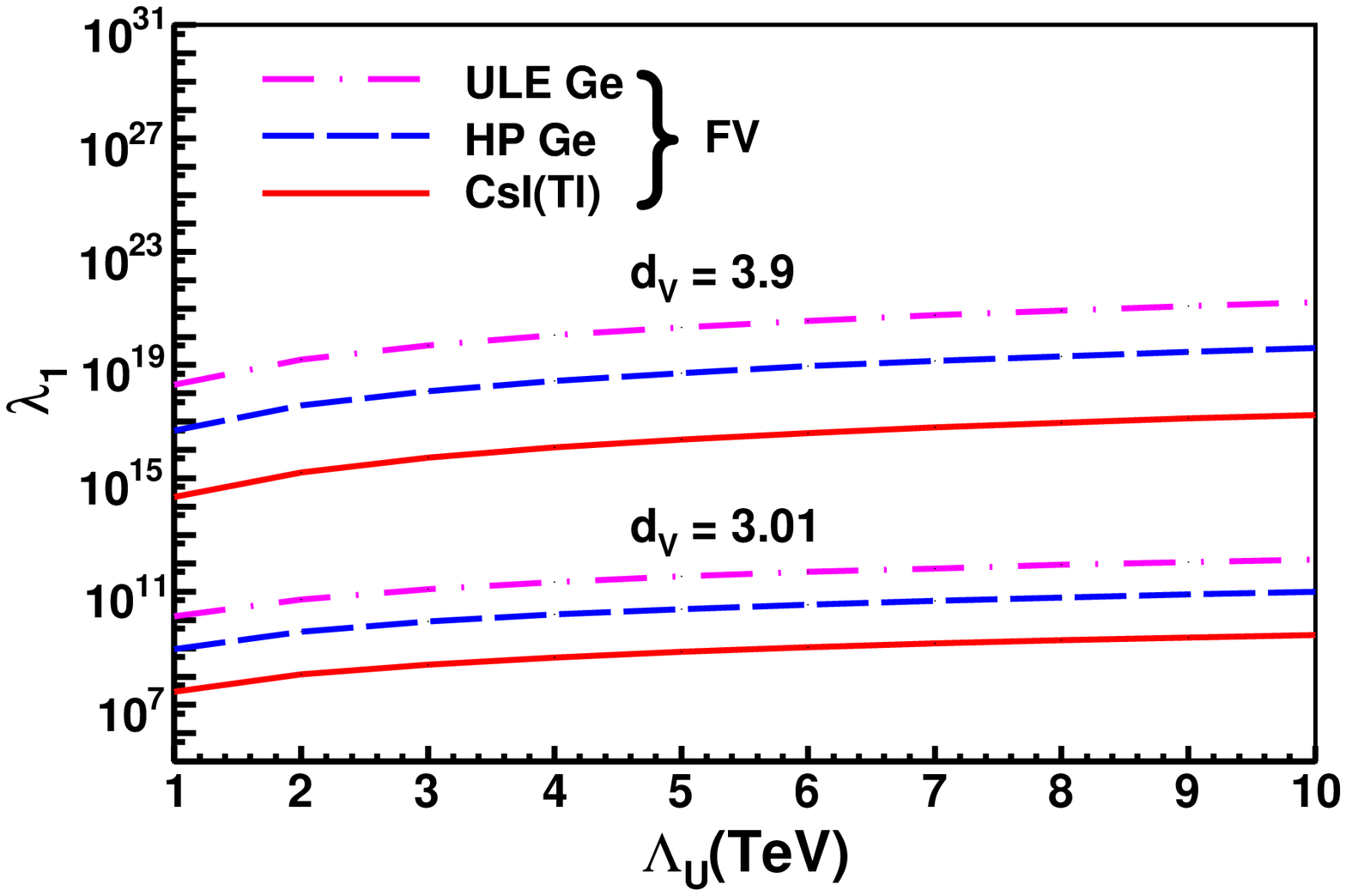} \\[2ex]
\includegraphics[width=8cm]{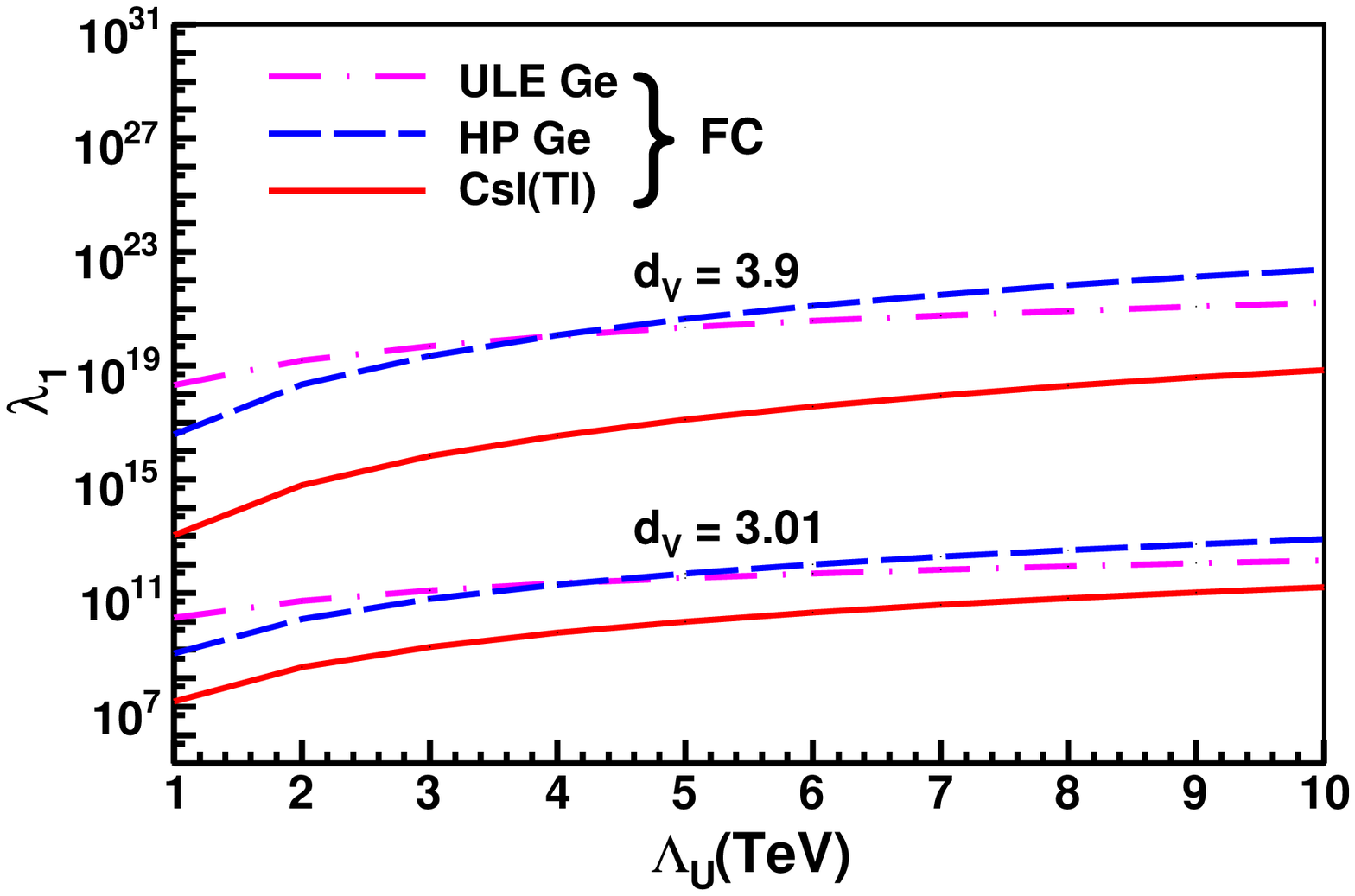}
\caption{ \label{upvector}
Constraints on UP with vector exchange $-$
Top: (a) The coupling  $\lambda_1$ versus $\dvec$
at $\Lambdaup = 1 ~ {\rm TeV}$.
The bounds apply for both FV and FC cases.
Middle (b) and Bottom (c):
Upper bounds on $\lambda_1$ at
different energy scales $\Lambdaup$
for FV and FC couplings, respectively,
at two values of $\dvec$.
Parameter space above the lines is excluded.
}
\end{figure}

\subsection{Non-Standard Neutrino Interaction}

The NSI parameters are constrained by the accuracy of the SM 
cross-section measurements. Accordingly, DS1-CsI(Tl) was adopted
for analysis. The NSI parameters of \Eq \ref{nsics} were the fitting
variables in the minimum-$\chi^2$ analysis.

Results from one-dimensional analysis are
presented in Table~\ref{tab:eelr}.
It can be inferred from Eq.~\ref{nsics} that the 
sensitivities for NU and FC couplings vary
as $\elr$ and $ [ \emlr ] ^2 ( [ \etlr ]^2 )$, respectively.
New limits on $\elr$, $\emlr$ and $\etlr$ were derived.
-he results on $\emlr$ and $\etlr$ are identical since 
their roles are symmetrical such that 
one-dimensional analysis cannot differentiate their effects.
The projected sensitivities due to a 
realistically achievable 2\% measurement of
the SM $\nuebar -$e cross-section 
with reactor neutrinos~\cite{texononue} are shown.
As comparison, we also list the constraints from
LSND $\nu_e -$e measurement~\cite{nsiboundlsnd} and those
from a combined analysis with data
from LEP, CHARM, LSND, and previous reactor
experiments~\cite{nsiboundcombined},
as well as a model-independent analysis on $\emlr$.

The allowed region at 90\% confidence level (CL) from two-parameter
analysis were displayed in Figures~\ref{nsitexono}a\&b in the $( \el ,
\er )$ and $( \etl , \etr )$ space, respectively, in which the
bounds from LSND~\cite{nsiboundlsnd} were overlaid. The
complementarity between the constraints due to $\nuebare$ 
and $\nuee$ scatterings can be readily seen. 
The constraints in the  $( \eml , \emr )$ plane 
are the same as those of  $( \etl , \etr )$ in 
Figure~\ref{nsitexono}b.

As comparison and for completeness, 
we also note the bounds
on NSI NU and FC couplings in the quark sector 
due to accelerator $\nu$N scattering experiments are
$| \epsilon^{q P}_{ee} | < 0.3 - 1$
and $| \epsilon^{q P}_{e \tau } | < 0.5 - 1.6$,
respectively~\cite{nsiboundlsnd}, where $P$ denotes 
the helicity-state $R/L$ while $q$=$u$ or $d$ quarks.
Projected sensitivities due to future experiments
on neutrino-nucleus coherent scatterings~\cite{nuNcohsca} are 
$|\epsilon_{ee}^{qL}+\epsilon_{ee}^{qR}|<0.001$ and 
$|\epsilon_{e\tau}^{qL}+\epsilon_{e\tau}^{qR}|<0.02$~\cite{nsinuN}.

\subsection{Unparticle Physics Parameters}

Since different ranges of the $\dsca$($\dvec$)
give different sensitivities to the cross-section,
all three data set were used in the UP analysis
for their complementarity.
The threshold value of DS2-HPGe was 
previously used to probe UP phenomenology
in Ref.~\cite{upbalantekin07}. 
A different cross-section formula was used 
and discussed in a later work~\cite{upbarranco09}.

Constraints on $\lambda_0$ at different 
$\dsca$ for scalar UP exchange
was derived at $\Lambdaup = 1 ~ {\rm TeV}$.
The results are shown in Figure~\ref{upscalar}a,
with bounds from the Borexino~\cite{upmontanino08} 
and MUNU~\cite{upbarranco09} experiments  superimposed.
The upper bounds for $\lambda_0$ at different
energy scale $\Lambdaup$ are shown in Figure~\ref{upscalar}b.
The data of DS2-HPGe 
provided better sensitivities at 
$\dsca < 1.3$,  while DS1-CsI(Tl) gave rise to more
stringent limits at larger $\dsca$.  

Constraints on vector UP couplings $\lambda_1$
as function of $\dvec$ are displayed in Figure~\ref{upvector}a.
Both FC and FV couplings give similar bounds 
in this parameter space.
The variations of $\lambda_1$ 
for the two cases are depicted in 
Figures~\ref{upvector}b\&c as 
function of the energy scale $\Lambdaup$.
The DS1-CsI(Tl) data set 
consistently provided more severe constraints
for vector UP exchanges, since
the couplings were enhanced at high energy
as indicated in Figure~\ref{dsdT}b.

Since $( d \sigma / d T )_{\cal U_S} \propto \lambda_0^4$
and $( d \sigma / d T )_{\cal U_V} \propto \lambda_1^4$
from Eqs.~\ref{cs:e-scalar}\&\ref{cs:e-vector}, respectively, 
the potentials on 
placing more severe constraints on the coupling constants 
due to improved experimental sensitivities are only modest.
When DS1-CsI(Tl) data would improve to 
provide a 2\% measurement of the SM cross-section, 
an improvement by a factor of 2 
to the sensitivities of $\lambda_0$ and $\lambda_1$ 
can be expected.
Similarly, the benchmark goals of
sub-keV ULEGe detectors for studying
neutrino-nucleus coherent scattering 
with reactor neutrinos are 
to achieve a background level of $\sim 1 ~ \cpkkd$
and ON$-$OFF subtraction of 1\%~\cite{nuNcohsca}.
If these are achieved, 
the sensitivities of $\lambda_0$ and $\lambda_1$ 
would enhance
by factor of 2 at $\dsca < 1.3$.

\section{Acknowledgments}

The authors appreciate discussions and comments from
A. B. Balantekin, M. Blennow, K. Cheung, S. Petcov, 
T. Rashba  and T. C. Yuan. 
This work is supported by contract 98-9628-M-001-013
under the National Science Council, Taiwan, and by contract 108T502
under TUBITAK, Turkey.

\end{document}